\documentclass[final,5p,times,twocolumn]{elsarticle}

\usepackage{amssymb}

 \usepackage{lineno}

\journal{J. of Atmospheric and Solar-Terrestrial Physics}

\begin{document}

\begin{frontmatter}



\title{Empirical evidence for a celestial origin of the climate oscillations and its implications}


\author{Nicola Scafetta $^{1,2}$}

 \address{$^{1}$Active Cavity Radiometer Irradiance Monitor (ACRIM) Lab, Coronado, CA 92118, USA}

 \address{$^{2}$Department of Physics, Duke University, Durham, NC 27708, USA.}

\begin{abstract}
  We investigate whether or not the decadal and multi-decadal climate oscillations have an astronomical origin.  Several global surface temperature records since 1850 and records deduced from the orbits of the planets present  very similar power spectra. Eleven frequencies with period between 5 and 100 years closely correspond in the two records. Among them, large climate oscillations with  peak-to-trough amplitude of about  0.1 $^oC$ and 0.25 $^oC$, and periods of about 20 and 60 years, respectively,  are synchronized  to the orbital periods  of Jupiter and Saturn. Schwabe and Hale solar cycles are also visible in the temperature records. A 9.1-year cycle is synchronized to the Moon's orbital cycles. A phenomenological model based on these astronomical cycles can be used to well reconstruct the temperature oscillations since 1850 and to make partial forecasts for the 21$^{st}$ century.  It is found that at least 60\% of the global warming observed since 1970   has been induced by the combined effect of the above  natural climate oscillations. The partial forecast indicates that climate may stabilize or cool until 2030-2040. Possible physical mechanisms are qualitatively discussed with an emphasis on the phenomenon of collective synchronization of coupled oscillators.\\ \\
  
  Please cite this article as: Scafetta, N., Empirical evidence for a celestial origin of the climate oscillations and its implications. \emph{Journal of Atmospheric and Solar-Terrestrial Physics} (2010), doi:10.1016/j.jastp.2010.04.015
\end{abstract}


\end{frontmatter}


\section{Introduction}

 Milankovic [1941] theorized  that variations in eccentricity, axial tilt, and precession of the  orbit of the Earth determine climate patterns such as  the 100,000 year ice age cycles of the Quaternary glaciation over the last few million years. The variation of the orbital parameters of the Earth is  due to the gravitational perturbations induced by the other planets of the solar system,  primarily Jupiter and Saturn.  Over a much  longer time scale the cosmic-ray flux record well correlates with the warm and ice periods of the Phanerozoic during the last 600 million years:  the cosmic-ray flux oscillations are likely due to the changing galactic environment of the solar system as it crosses the spiral arms of the Milky Way [Shaviv, 2003, 2008; Shaviv and Veizer, 2003; Svensmark, 2007]. Over  millennial and secular time scales several authors have found that variations in total solar irradiance  and variations in solar modulated cosmic-ray flux  well correlate with climate changes: see for example: Eddy, 1976; Hoyt  and  Schatten, 1997; White \emph{et al.}, 1997; van Loon and Labitzke, 2000; Bond \emph{et al.}, 2001; Kerr, 2001; Douglass and Clader, 2002; Kirkby, 2007; Scafetta and West, 2005, 2007, 2008; Shaviv, 2008; Eichler \emph{et al.}, 2009; Soon, 2009; Meehl \emph{et. al.}, 2009; Scafetta 2009, 2010. Also the annual cycle has an evident astronomical origin.

The above results suggest that the dominant drivers of the climate oscillations have a celestial origin. Therefore, it is legitimate to investigate whether the climate oscillations with a time scale between 1 and 100 years, can be interpreted in astronomical terms too.

Global surface temperature has risen [Brohan \emph{et al.}, 2006] by about 0.8 $^oC$ and 0.5 $^oC$ since 1900 and 1970, respectively. Humans may have partially contributed to this global warming through greenhouse gas (GHG) emissions [IPCC, 2007]. For instance, the IPCC claims that more than 90\% of the observed warming since 1900 and practically 100\% of the observed warming since 1970 have had  an anthropogenic origin (see figure 9.5 in  IPCC, AR4-WG1). The latter conclusion derives merely from the fact that climate models referenced by the IPCC cannot explain the warming occurred since 1970 with any known natural mechanism. Therefore, several scientists have concluded that this warming has been caused by anthropogenic GHG emissions that greatly increased during this same period. This theory is known as the \emph{anthropogenic global warming theory} (AGWT).

However,  the anthropogenic GHG emissions have increased monotonically since 1850 while the global temperature record did not. Several oscillations are seen in the data since 1850, including a global cooling since 2002: see Figure 1. If these climate oscillations are natural, for example induced by astronomical oscillations, they would determine how climate change should be interpreted [Keenlyside \emph{et al.}, 2008]. In fact,  during its cooling phase a natural multi-decadal oscillation can hide a global warming caused by human GHG emissions or, alternatively, during its warming phase a natural oscillation can accentuate the warming. If the natural oscillations of the climate are not properly recognized and taken into account, important climate patterns, for example the global warming observed from 1970 to 2000, can be erroneously interpreted. Indeed, part of the 1970-2000 warming could have been induced by a multi-decadal natural cycle during its warming phase that the climate models used by the IPCC have not reproduced.

The IPCC [2007] claims that the climate oscillations are induced by some still poorly understood and modeled internal dynamics of the climate system, such as the ocean dynamics. However, the oscillations of the atmosphere and of the ocean, such as the Pacific Decadal Oscillation (PDO) and the Atlantic Multidecadal Oscillation (AMO), may be induced by complex extraterrestrial periodic forcings that are acting on the climate system in multiple ways. Indeed, the climate system is characterized by interesting cyclical patterns that remind astronomical cycles.

\begin{figure}[t!]
\includegraphics[angle=-90,width=21pc]{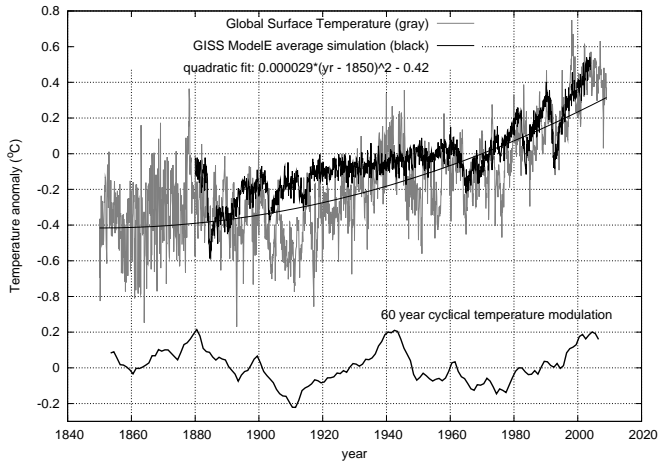}
\caption{ Top: Global surface temperature anomaly (gray) [Brohan \emph{et al.}, 2006] against the GISS ModelE average simulation (blak) [Hansen \emph{et al.}, 2007]. The  figure also shows the quadratic upward trend of the temperature. Bottom:  an eight year moving average smooth of the temperature detrended of its upward quadratic trend. This smooth reveals a quasi-60 year modulation. }
\end{figure}

For example, surface temperature records are characterized by decadal and bi-decadal oscillations which are usually found in good correlation with the (11-year) Schwabe and the (22 year) Hale solar cycles [Hoyt and Schatten, 1997; Scafetta and West, 2005; Scafetta, 2009]. However, longer cycles are of interest herein.

 Klyashtorin and Lyubushin [2007] and Klyashtorin et al. [2009]  observed that several centuries of climate records (ice core sample, pine tree samples, sardine and anchovy sediment core samples, global surface temperature records, atmospheric circulation index, length of the day index, fish catching  productivity records, etc.) are characterized by  large 50-70 year and 30-year periodic cycles.  The quasi-60 year periodicity has been also found  in secular  monsoon rainfall records from India, in proxies of monsoon rainfall from Arabian Sea sediments and in rainfall over east China [for example see the following works and their references: Agnihotri \emph{et al.}, 2002; Sinha A. \emph{et al.} (2005); Goswami \emph{et al.}, 2006; Yadava  and Ramesh 2007]. Thus, several records indicate that the climate is characterized by a large quasi-60 year periodicity, plus larger secular climatic cycles and smaller decadal cycles. All these cycles cannot be explained with anthropogenic emissions. Errors in the data, other superimposed patterns (for example, volcano effects and longer and shorter cycles) and some chaotic pattern in the dynamics of these signals may sometimes mask the 60-year cycle.

A multi-secular climatic record that shows a clear quasi-60 year oscillation is depicted in Figure 2: the G. Bulloides abundance variation record found in the Cariaco Basis sediments in the Caribbean sea since 1650 [Black \emph{et al.}, 1999]. This record is an indicator of the trade wind strength in the tropical Atlantic ocean and of the north Atlantic ocean atmosphere variability. This record  shows five 60-year large cycles. These cycles correlate well with the 60-year modulation of the global temperature observed since 1850 (the correlation is negative). On longer time scales, periods of high G. Bulloides abundance correlate well with periods of reduced solar output (the well-known Maunder, Sp\"orer, and Wolf minima), suggesting a solar forcing origin of these cycles [Black \emph{et al.}, 1999].

\begin{figure}[t!]
\includegraphics[angle=-90,width=21pc]{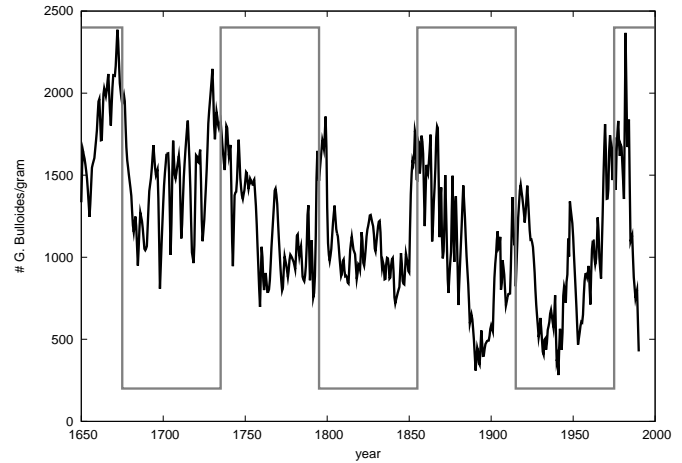}
\caption{ Record of G. Bulloides abundance variations (1-mm intervals)   from  1650 to 1990 A.D. (black
line) [Black \emph{et al.}, 1999]. The gray vertical lines highlight 60-year intervals. Five quasi-60 year cycles are seen in this record, which is a proxy for the Atlantic variability since 1650.}
\end{figure}

 Patterson et al. (2004) found 60-62 year cycles in sediments and cosmogenic nuclide records in the NE Pacific. Komitov (2009) found similar cycles in the number of the middle latitude auroras from 1700 to 1900.
 A cycle of about 60 years has been detected in the number of historically recorded meteorite falls in China from AD 619 to 1943 and in the number of witnessed falls in the world  from 1800 to 1974 [Yu \emph{et al.}, 1983]. Ogurtsov \emph{et al.} [2002] found   a 60-64 year cycle in $^{10}Be$, $^{14}C$ and Wolf number over the past 1000 years.  The existence of a 60-year signal has been found in the Earth's angular velocity and in the geomagnetic field  [Roberts \emph{et al.}, 2007]. These results clearly suggest an astronomical origin of the  60-year variability found in several climatic records.

 Interestingly,  the traditional Chinese calendar, whose origins can be traced as far back as the 14$^{th}$ century BCE, is arranged in major 60-year cycles [Aslaksen, 1999]. Each year is assigned a name consisting of two components. The first component is one of the 10 \emph{Heavenly Stems}  (\emph{Jia, Yi, Bing}, etc.), while the second component is one of the 12 \emph{Earthly Branches}  that features the names of 12 animals (\emph{Zi, Chou, Yin}, etc.). Every 60 years the stem-branch cycle repeats. Perhaps, this sexagenary cyclical calendar was inspired by climatic and astronomical observations.

Some studies [Jose, 1965; Landscheidt, 1988, 1999; Charv\'atov\'a, 1990, 2009;  Charv\'atov\'a and St\v{r}e\v{s}t\'ik, 2004; Mackey, 2007; Wilson \emph{et al.}, 2008; Hung, 2007]   suggested that solar variation may be driven by the  planets through  gravitational  spin-orbit coupling mechanisms and tides. These authors have used the inertial motion of the Sun around the center of mass of the solar system (CMSS) as a proxy for describing this phenomenon.  Then, a varying Sun would influence the climate by means of several and complicated mechanisms and feedbacks [Idso and Singer, 2009]. Indeed,  tidal patterns on the Sun well correlate with large solar flare occurrences, and the alignment of Venus, Earth and Jupiter well synchronizes with the 11-year Schwabe solar cycle [Hung, 2007].

In addition, the Earth-Moon system and the Earth's orbital parameters can also be directly modulated by the planetary oscillating gravitational and magnetic fields, and synchronize with their frequencies [Scafetta, 2010]. The Moon can influence the Earth through gravitational tides and orbital oscillations [Keeling and Whorf, 1997, 2000; Munk and Wunsch, 1998;  Munk  and Bills, 2007].

 It could be argued that planetary tidal forces are \emph{weak} and unlikely have any physical outcome. It can also be argued that the tidal forces generated by the terrestrial planets are comparable or even larger than those induced by the massive jovian planets.  However, this is not a valid physical rationale because still little is  known about the solar dynamics and the terrestrial climate. Complex systems are usually characterized by feedback mechanisms that can   amplify the effects of weak periodic forcings also by means of resonance and collective synchronization processes [Kuramoto, 1984; Strogatz, 2009].  Thus, unless the physics of a system is not clearly understood, good empirical correlations at multiple time scales cannot be dismissed just because the microscopic physical mechanisms may be still obscure and need to be investigated.

The above theory implies the existence of direct and/or indirect links between the motion of the planets and the climate oscillations, essentially claiming that the climate is synchronized to the natural oscillations of the solar system, which are driven  by the movements of the planets around the Sun.  If this theory is correct, it can be efficiently used for interpreting climate changes and forecasting climate variability because the motion of the planets can be rigorously calculated.

In this paper we investigate this theory by testing a synchronization hypothesis, that is, whether the planetary motion and the climate present a common set of frequencies. Further, we compare the statistical performance of a phenomenological planetary model for interpreting the climate oscillations with that of a typical major general circulation model adopted by the IPCC. Our findings show that a planetary-based climate model would largely outperform the traditional one in reconstructing the oscillations observed in the climate records.

\section{Climate and planetary data and their spectral analysis}

Figure 1 shows the global surface temperature (HadCRUT3) [Brohan \emph{et al.}, 2006] from 1850 to 2009 (monthly sampled) against an advanced general circulation model average simulation   [Hansen et al., 2007]. This general circulation climate model uses all known climate forcings and all known climate mechanisms. This is one of the major general circulation climate models adopted by the IPCC [2007]: this model attempts to reconstruct more than 120 years of climate.

 The temperature record presents a clear 60-year cycle that oscillates around an upward trend. In fact, we see the following 30-year trends: 1850-1880, warming; 1880-1910, cooling; 1910-1940, warming; 1940-1970, cooling; 1970-2000, warming; and, therefore, a probable cooling from 2000 to 2030. This 60-year modulation has a peak-to-trough amplitude of about 0.3-0.4 $^oC$, as shown at the bottom of Figure 1. Figure 2 suggests that a quasi-60 year cycle is present in the climate system since at least 1650.

\begin{figure} [t!]
\includegraphics[angle=-90,width=21pc]{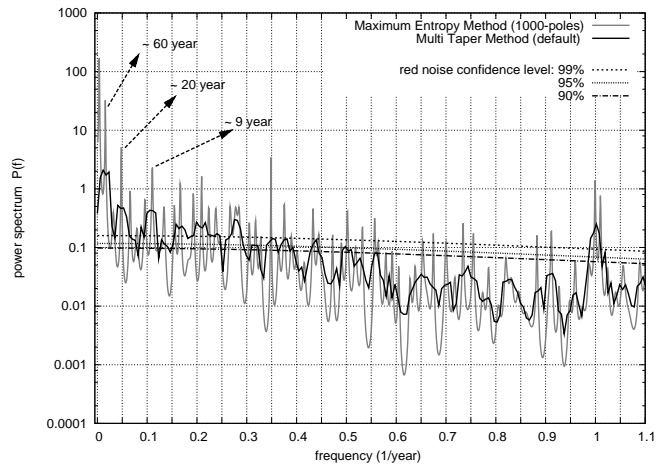}
\caption{Power spectrum [Ghil \emph{et al.}, 2002] estimates of the global temperature. Two methods are adopted: the maximum entropy method (1000 poles) and the multi-taper method with three confidence levels relative to the null hypothesis of red noise. (In the multi-taper analysis  the quadratic trend of the temperature is removed for statistical stationarity).}
\end{figure}

On the contrary, the general circulation climate model simulation presents an almost monotonic warming trend that follows the monotonic upward trend of the greenhouse gas concentration ($CO_2$ and $CH_4$) records. The model simulation does not appear to fit the temperature record before 1970. Indeed, it fails to model the large 60-year temperature cyclical modulation. This failure is also true for the IPCC [2007] multi-model global average surface temperature during the 20$^{th}$ century (see the IPCC's figure SPM.5): a fact that indicates that this failure is likely common to all models adopted by the IPCC [Note \#1]. The climate model simulation is slightly modulated by the aerosol record, and is interrupted by sudden volcano eruptions that cause a momentary intense cooling. The major volcano eruptions are: Krakatau (1883), several small eruptions (in the 1890s), Santa Mar\'ia (1902),  Agung (1963), Awu (1968),   El Chich\'on (1982) and Pinatubo (1991).

Figure 3 shows the power spectrum [Ghil \emph{et al.}, 2002] of the global surface temperature (monthly sampled). Two methods are adopted: the maximum entropy method (1000 poles) and the multi-taper method against the null hypothesis of red noise with three confidence levels. The figure shows strong peaks at 9, 20 and 60 years with a 99\% confidence against red noise background. The graph also shows a clear annual cycle ($f=1$ $yr^{-1}$) and several other cycles with a 99\% confidence. The harmonic signal Fisher (F) test gives a significance level larger than 99\% and 95\% for the 60 and 20 year cycles, respectively. The Blackman-Tukey correlogram produces a spectrum equivalent to that obtained with the maximum entropy method, but its peaks are less sharp and do not have a good resolution. In the following, the maximum entropy method is used because with a proper number of poles it better resolves the low frequency band of the spectrum and produces very sharp peaks [Priestly, 2001].

\begin{figure}[t!]
\includegraphics[angle=-90,width=21pc]{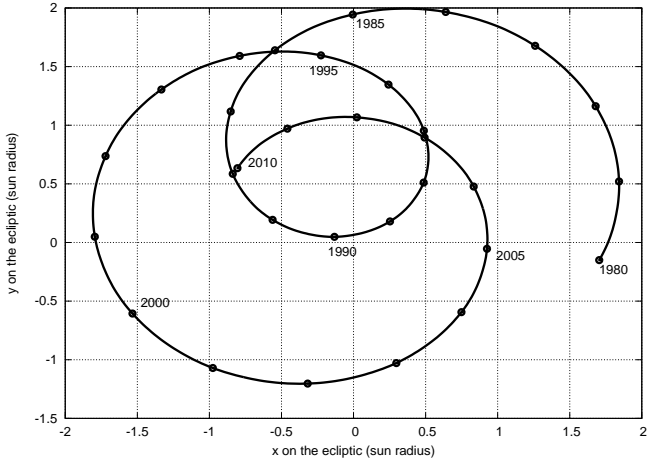}
\caption{A projection of the solar orbit relative to the CMSS on the ecliptic plane from 1980 to 2010.}
\end{figure}

Determining at least the major planetary cycles is simple. Jupiter's period is 11.86 years while Saturn's period is 29.42 years. Thus, the following five major cycles are present: about 10 years, the opposition-synodic period of Jupiter and Saturn; about 12 years, the period of Jupiter;  about 20 years, the synodic period of Jupiter and Saturn (synodic period $=(P_1^{-1}-P_2^{-1})^{-1}$ with $P_1<P_2$, $P_1$ and $P_2$ are the periods of the two planets); about 30 years, the  period of  Saturn; about 60 years, the repetition of the combined orbits of Jupiter and Saturn. Moreover, there is the 11-year Schwabe solar cycle that can be produced by the alignment of Venus, Earth and Jupiter [Hung, 2007], and the 22-year Hale magnetic solar cycle that is made of two consecutive Schwabe cycles. Uranus (84-year period) and Neptune (164.8-year period) together with Jupiter and Saturn regulate secular, bi-secular and millennial cycles, which are observed in the radiocarbon and sunspot records [Jose, 1965; Suess, 1980; Ogurtsov et al., 2002]. Some orbital combinations of the terrestrial and jovian planets can induce other minor cycles. In conclusion, the major solar system oscillations within the secular scale have a period of about 10-11, 12, 20-22, 30, 60 years with a 5\% error that is due to the elliptical shape of the planetary orbits.

\begin{figure}[t!]
\includegraphics[angle=0,width=21pc]{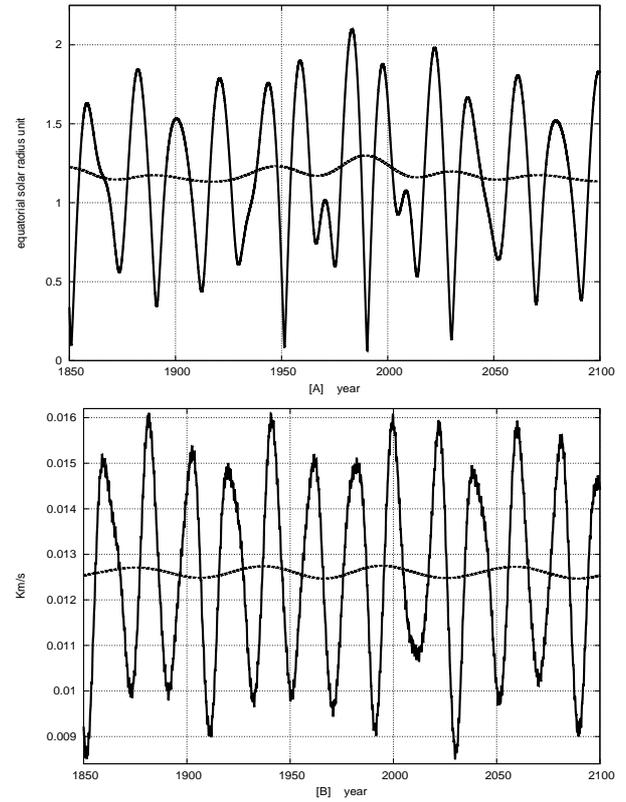}
\caption{[A] Distance and [B] speed of the Sun relative to the CMSS. Note the 20 and 60 year oscillations (smooth dash curves), which are due to the orbits of Jupiter and Saturn. In addition, a longer cycle of about 170-180 years is clearly visible in [A]. This is due to the additional influence of Uranus and Neptune. }
\end{figure}

The idea proposed here is that the climate oscillations are described by a given, even if still unknown, physical function that depends on the orbits of the planets and their positions. We observe that all major planetary cycles  can be determine by choosing nearly any function of the planetary orbits. In fact, different physical records of the same orbits present a similar set of frequency peaks, although the relative amplitude of the spectral peaks can differ from function to function. A simple analogy can be found in music where different instruments can play the same music. Because the instruments are different, the timber of their sound is different, but because these instruments are playing the same notes, the frequencies of the produced music are the same. Therefore, to determine the frequency peaks, any function of those frequencies can be analyzed.  For example, a set of planetary frequencies can be determined by using as proxy the distance of the Sun about the center of mass of the solar system (CMSS), or it is possible to choose its speed (SCMSS).

\begin{figure}[t!]
\includegraphics[angle=0,width=21pc]{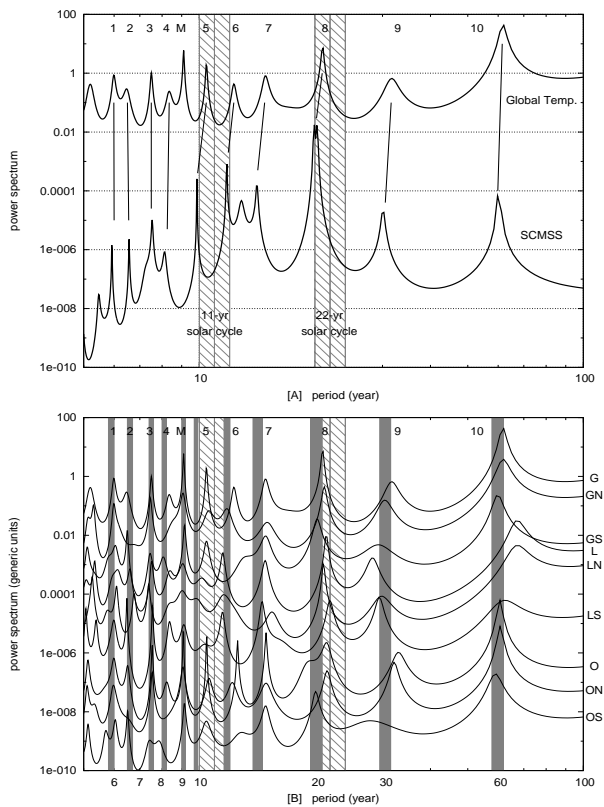}
\caption{Maximum entropy power spectrum analysis against the period: [A] Global temperature (top) and SCMSS (bottom) from 1850 to 2009; [B] Power spectra of global (G), N. Hemisphere (GN), S. Hemisphere (GS), land (L), N. Hemisphere land (LN), S. Hemisphere land (LS), ocean (O), N. Hemisphere ocean (ON) and S. Hemisphere ocean (OS). The SCMSS frequencies are represented with gray filled boxes. Two gray dash box indicate the  $11\pm 1$  and $22\pm2$ year known solar cycles. A gray box at 9.1-year corresponds to a lunar cycle. }
\end{figure}

The orbit of the Sun around the center of mass on the solar system can be easily evaluated using the NASA Jet Propulsion Laboratory Developmental Ephemeris. Figure 4 shows a projection of this orbit on the ecliptic plane from 1980 to 2010. Several physical variables can be evaluated from this complex orbit such as the distance of the Sun from the CMSS and its speed, SCMSS. Figure 5A and 5B show these two curves for a few centuries, respectively.

Irregular cycles with an average period of about 20 years are clearly visible in Figure 5. These cycles are determined by the synodic period of Jupiter and Saturn, as explained above. A 60-year cycle is also clearly visible in the figure in the smooth dash curves. A longer secular cycle of about 178 years [Jose, 1965], which is mostly determined by the synodic period of Uranus and Neptune (about 171.4 years), is also present and evident in Figure 5A in the secular modulation.
Several other shorter cycles are present, but not easily visible in these records.

Herein, the CMSS speed   (SCMSS) (Figure 5B) is preferred as a convenient sequence for estimating the frequencies of the planets' orbits within the secular scale.  The SCMSS record was not used in  previous publications. This record   stresses a set of frequencies that can be more directly associated with the low frequency solar disturbances induced by Jupiter and Saturn within the secular scale.

Figure 6A compares the power spectra (by using the maximum entropy method, 1000 poles) of the global temperature record and of the SCMSS record against the period (not the frequency as done in Figure 3) for visual convenience. In Figure 6B the same analysis is applied to nine global temperature records (global, land and ocean for both northern and southern hemispheres [Brohan \emph{et. al.}, 2006])  from Jan-1850 to Jan-2009. The power spectra   of the nine temperature records (with an arbitrary vertical shift, black curves) are plotted against the frequency bands (gray bands) of the SCMSS record shown in Figure 6A. The peaks shown in the two figures all have a statistical significance above 99\%, as those shown in Figure 3. The periods of these peaks are reported in Table 1. We added two gray dash boxes to indicate the $11\pm 1$ and $22\pm2$ year solar cycles and a gray box at 9.1 years that, as explained in the next section, appears to be linked to a lunar cycle. The frequency peaks are numbered for visual convenience.

\begin{figure} [t!]
\includegraphics[angle=-90,width=21pc]{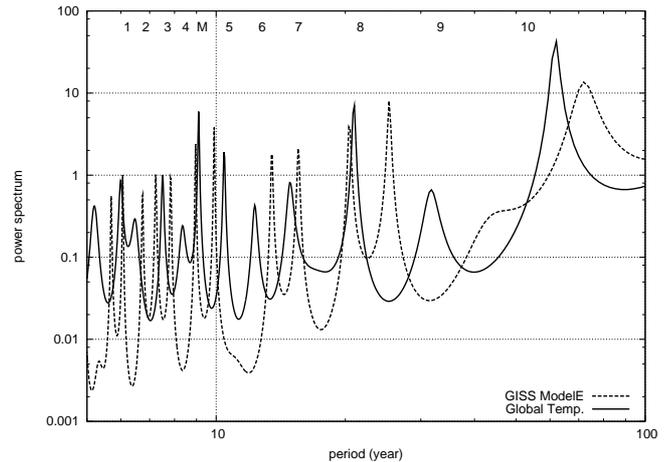}
\caption{Comparison between power spectra of the global temperature (solid) shown in Figure 6 and of the GISS ModelE average simulation (dash) depicted in Figure 1. (Maximum entropy method 1000 poles). The periods of the eleven peaks are reported in Table 2. It is evident from the figure that several peaks do not correspond in the two records.}
\end{figure}

 The power spectra of the temperature records appear quite similar to each other. This indicates coupling and synchronization among all terrestrial regions and/or the possibility that all regions of the Earth are forced by a common external forcing. Ten on 11 cycles with periods within the range of 5-100 years, as indicated in the figure, appear to be reproduced by the power spectrum of SCMSS that reproduces the major planetary frequencies. The temperature cycles \#5 and \#8 are also compatible with the $11\pm 1$ and $22\pm2$ year Schwabe and Hale solar cycles, respectively.

There are slight differences among the correspondent frequency peaks of the temperature records.  However, for physical reasons, the northern and southern hemispheres, as well as the ocean and the land regions, must be mutually synchronized because of their mutual physical couplings. Thus, these records should present the same decadal and multidecadal frequencies. The discrepancies among the peaks can be due to non-linear fluctuations around limit cycles, as often happens in chaotic systems, or to some errors in the temperature data. Small discrepancies can also be due to the regression models implemented in the maximum entropy method [Priestly, 2001].  About possible errors in the temperature records we notice that: 1) the temperature data  before 1880 may be  less accurate  than the data after 1880 [Brohan \emph{et. al.}, 2006];     2) there may be some additional errors in some region and during some specific  period [Thompson \emph{et al.}, 2008]; 3) the urban heat island effects may have been poorly corrected in some regions [McKitrick and  Michaels, 2007; McKitrick, 2010]. Random oscillations and/or possible errors in the temperature data can slightly shift the spectral peaks observed in the data from their true values.

The existence of  major errors in some temperature records appears reasonable from the power spectral analysis alone. In fact,   the northern land record (LN) presents a cycle of about 67.4 year period, while the southern land record (LS) as well as the northern and the southern ocean records (ON and OS) present  cycles with a clearer 60-year periodicity. It is unlikely that the northern land temperature is thermodynamically decoupled from the rest of the world. Probably the northern land temperature record is skewed upward by uncorrected urban heat island effect [McKitrick and  Michaels, 2007; McKitrick, 2010], or it  contains  some other errors. Because of these errors, the spectral analysis of the northern land temperature record gives a slightly longer cycle.

To test whether the matching found between the temperature records and the celestial records is just coincidental, in Figure 7 we compare the power spectrum of the global temperature with the power spectrum of the GISS ModelE average simulation of the global surface temperature depicted in Figure 1. Figure 7 shows that several temperature cycles are not reproduced by the GISS ModelE average simulation. Among the cycles that the model fails to reproduce there is the large and important 60-year modulation. Table 2 reports the value of the 11 periods found in the GISS ModelE average temperature simulation.

\begin{figure}[t!]
\includegraphics[angle=-90,width=21pc]{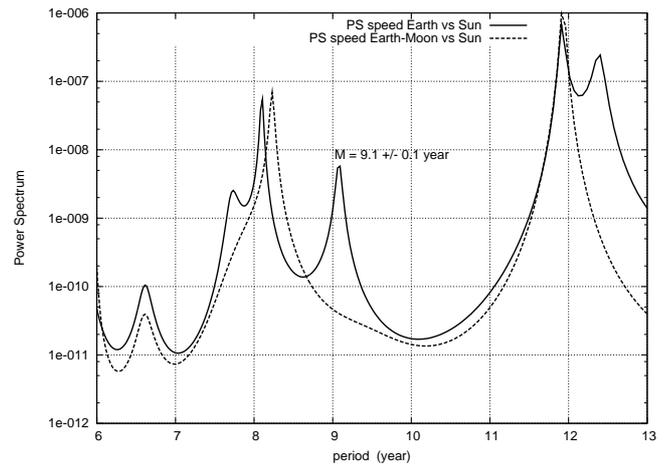}
\caption{Maximum entropy method (900 poles) power spectra of the speed of the Earth relative to the Sun (not to the CMSS) (solid line) and of the speed of the center of mass of the Earth-Moon system relative to the Sun (dash line). Note the peak `M' at 9.1 years that is present only in the speed of the Earth relative to the Sun. This result proves that the cycle `M' at 9.1 years is caused by the Moon orbiting the Earth.}
\end{figure}

\section{The lunar origin of the 9.1-year temperature cycle}

The nine temperature records show a strong spectral peak at 9-9.2 years (cycle `M' in Figure 6). This cycle is absent in the SCMSS power spectrum. This periodicity is exactly between the period of the recession of the line of lunar apsides, about 8.85 years, and half of the period of precession of the luni-solar nodes, about 9.3 years (the luni-solar nodal cycle is 18.6 years). Thus, this 9.1-year temperature cycle can be induced by long lunar tidal cycles. In fact, this 9.1-year cycle is particularly evident in the ocean records (Figure 6B), and the oceans  are quite sensitive to the lunar gravitational tides.

Indeed, there are many studies that have suggested a possible influence of the Moon upon climate [Keeling and Whorf, 1997, 2000; Munk and Wunsch, 1998; Ramos da Silva and Avissar, 2005;  Munk  and Bills, 2007; McKinnell and  Crawford, 2007]. After all, the phenomenon of lunar tides and their cycles are well known and clearly present in the ocean records. Thus, the Moon may alter climate by partially modulating the ocean currents via gravitational forces through  its long-term lunar tidal cycles.

It is possible to prove that the 9.1-year temperature cycle has a lunar origin. By using the NASA Jet Propulsion Laboratory Developmental Ephemeris two records from 1850 to 2009 are obtained:  the speed of the Earth relative to the Sun and the speed of the center of mass of the Earth-Moon system relative to the Sun. It is evident that the only difference between the two records is that  the former contains an additional small modulation due to the orbit of the Earth around the center of mass of the Earth-Moon system. This small modulation is only due to the presence of the Moon. Figure 8 shows the power spectra of the two records.   As the figure shows, the speed of the Earth relative to the Sun presents a clear frequency peak at $9.1\pm0.1$ years, which is missing in the speed of the center of mass of the Earth-Moon system.  This fact  indicates that the strong peak around 9.1 years found in the temperature records is due to the presence of the Moon and its long-term orbital cycles around the Earth.

 Figure 8  indicates that the temperature cycles \#1 (at 6 years), \#2 (at 6.6 year), \#4 (at 8.2 years) and \#6 (at 12 years) are also found in the speed of the Earth relative to the Sun. Other longer temperature cycles
[for example: \#7 (at 15 years), \#8 (at 20 years), \#9 (at 30 years) and \#10 (at 60 years)] are present in these speed records (not shown in the figure). This fact indicates that the planets slightly perturb the Earth-Moon system, as it is physically evident because their  gravitational  forces are acting on the Earth-Moon system too.

\begin{figure}[t!]
\includegraphics[angle=0,width=21pc]{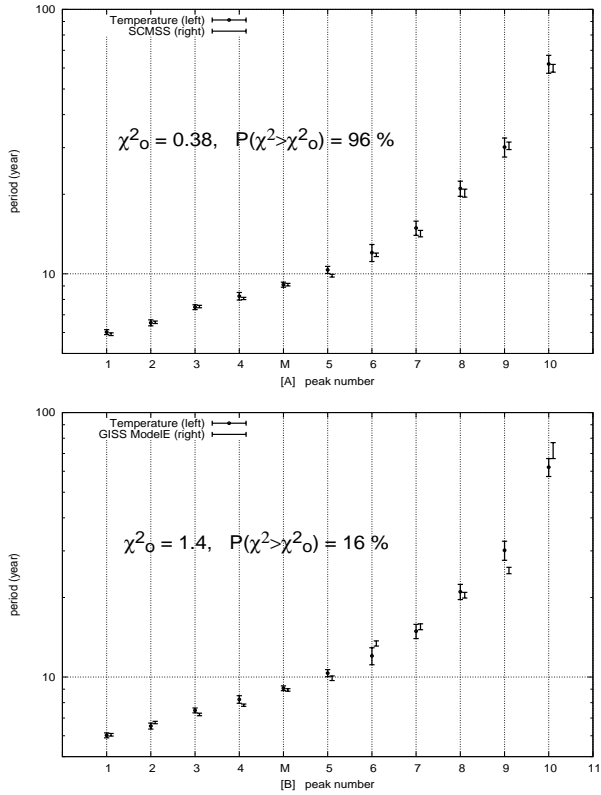}
\caption{[A] Coherence test between the average periods of the eleven cycles in the temperature records (left) and the ten cycles in the SCMSS (right) plus the cycle `M' at 9.1-year cycle associated to the Moon from Figure 8.  [B] Coherence test between the average periods of the eleven cycles in the temperature records (left) and the 11 cycles found in the GISS ModelE simulation in Figure 9 (right).  The figures depict the data reported in Table 2.  }
\end{figure}

\section{Analysis of the coherence}

Because the temperature cycles appear to fluctuate around ideal limit cycles, we evaluate the average value of each of the 11 peaks depicted in Figure 6 using the nine temperature records as reported in Table 1. Each average value can be interpreted as the best estimate of the limit cycles around which the temperature, at that specific frequency band, oscillates. We compare this set of temperature limit cycles against the cycles of the SCMSS record, and estimate the coherence between the two sets: see Table 2 for details.

Figure 9 shows the results of the coherence analysis. This figure also tests which model may better reconstruct the temperature oscillations: one based on the orbits of the planets, the Sun and the Moon, or a traditional IPCC's general circulation climate model that ignores any complex astronomical-climate link, such as the GISS ModelE.

 Figure 9A compares the 11 temperature and the SCMSS cycles plus the lunar 9.1-year cycle found in Figure 8. Figure 9A shows that there exists a remarkable correspondence between the two sets of frequency within the uncertainty. The comparison of the two sets gives a reduced $\tilde{\chi}_O^2=0.38$ (with 11 degrees of freedom). This indicates that the planetary orbits are significantly coherent to the climate oscillations [$P_{11}(\tilde{\chi}^2 \geq \tilde{\chi}_O^2)\approx 96\%$].

  Note that the temperature cycle \#5 ($10.35\pm0.3$ yr) is between the   SCMSS cycles \#5 ($9.84\pm0.12$ yr) and the $11\pm1$ year sunspot cycle, and the temperature cycle \#8 ($21\pm1.4$ yr) is between the   SCMSS cycles \#8 ($20.2\pm0.7$ yr) and the $22\pm2$ year Hale cycle.  If we substitute the SCMSS cycles \#5 and \#8 with the  $11\pm 1$  and $22\pm2$  year solar cycles the comparison of the two sets of frequencies gives a reduced $\tilde{\chi}_O^2=0.21$ [$P_{11}(\tilde{\chi}^2 \geq \tilde{\chi}_O^2)\approx 100\%$]. Thus, the association between the climate cycles and the major celestial cycles of the Sun, the planets and the Moon is statistically extremely significant.

The same cannot be said when the power spectrum of the temperature records is compared against the power spectrum of the GISS ModelE average simulation: see Figures 7 and 9B. At least seven on eleven peaks are statistically incompatible between the two records because of their $\chi_O^2>1$. These are the cycles \#2, 3, 4, 5, 6, 9 and 10. Thus, the climate model does not reproduce even the large 60-year cycle observed in the temperature records. From Table 2, the comparison between the temperature and the GISS ModelE simulation cycle sets gives a reduced $\tilde{\chi}_O^2 \approx 1.4$. In this case $P_{11}(\tilde{\chi}^2 \geq \tilde{\chi}_O^2)\approx 16\%$. The latter value is significantly lower than the 96\% value found above and depicted in Figure 9A. This indicates that a planetary-based climate model would largely outperform traditional general circulation models, such as the GISS ModelE, in reconstructing the oscillations observed in the climate records.

Note that the power spectrum of the GISS ModelE average simulation presents peaks at 10 and 20 years. These peaks appear close to the temperature peaks \#5 and \#8. However, in the model simulation these peaks are likely due to some regularity found in the volcano signal [North and Stevens, 1998]. For example, a decade separates the eruptions of the El Chich\'on (March, 1982) and of the Pinatubo (June, 1991). However,  volcano activity should not be considered as the true cause of these temperature decadal and bi-decadal cycles because these temperature cycles have been found to be in phase and positively correlated with the  solar induced cycles [Scafetta and West, 2005; Scafetta, 2009].

\begin{figure}[t!]
\includegraphics[angle=0,width=21pc]{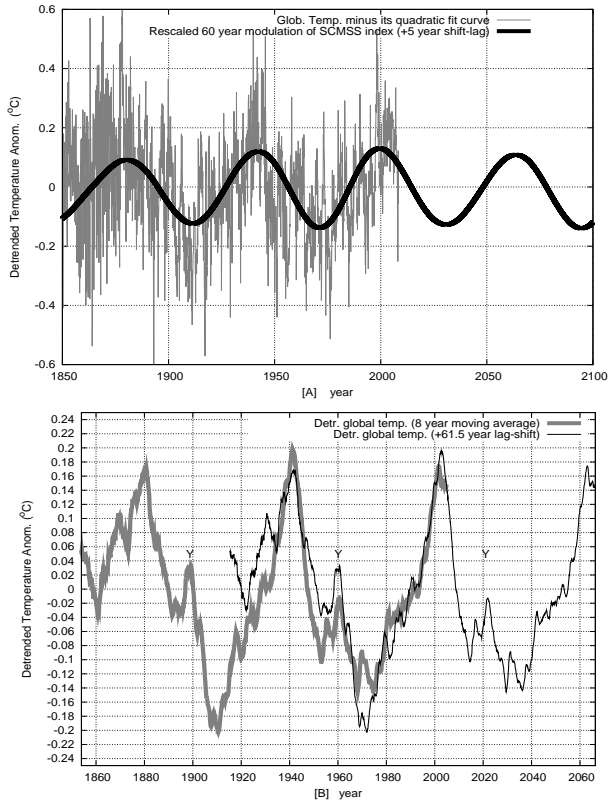}
\caption{[A] Rescaled SCMSS 60 year cycle (black curve) against  the global surface temperature record (grey) detrended of its quadratic fit; [B] Eight year moving average of the global temperature detrended of its quadratic fit and plotted against itself shifted by 61.5 years. Note the perfect correspondence between the 1880-1940 and 1940-2000 periods.  Also a smaller cycle, whose peaks are indicated by the letter ``Y'', is clearly visible in the two records. This smaller cycle is mostly related to the 30-year modulation of the temperature. These results reveal the natural origin of a large 60-year modulation in the temperature records.}
\end{figure}

In conclusion, Figures 9A and 9B imply that a model based on celestial cycles would reproduce the temperature oscillations much better than  typical general circulation models such as those adopted by the IPCC. The IPCC models  do not contain  any complex astronomical forcings nor complex feedback mechanisms that would amplify a solar input on climate, such as a cloud modulation via cosmic ray flux variation [Kirkby, 2007; Svensmark \emph{et al.}, 2009].

The good coherence between the celestial and the temperature records indicates that the two records share a compatible physical \emph{information} [Scafetta and West, 2008]. The Earth's climate just looks synchronized to the astronomical oscillations of the solar system.

The failure of the climate models, which use all known climate forcing and mechanisms, to reproduce the temperature oscillations at multiple time scales, including the large 60-year temperature modulation,  indicates that the current climate models are missing fundamental climate mechanisms. The above findings indicate, with a very high statistical confidence level, that major climate forcings have an astronomical origin and that these forcings are not included in the current climate models.

\section{Reconstruction and forecast of the climate oscillations}

 Herein, we reconstruct the oscillations of the climate with the large 20 and 60-year astronomical oscillations.  Reconstructing smaller time scales is possible but it requires more advanced mathematical techniques: for instance it would require the determination of the correct phase of the cycles and their exact frequencies. This more advanced reconstruction is left to another study.   Because the temperature appears to be growing since 1850 with at least an accelerating rate, we can approximately detrend this upward trend from the global temperature data  using a quadratic fit function:  $y=2.8*10^{-5}(x-1850)^2-0.41$.

Figure 10A shows the global surface temperature record detrended of the its quadratic fit. Two large and clear sinusoidal-like cycles with a 60-year period and with a peak-to-trough amplitude of about 0.25 $^oC$ appear. This temperature oscillation is compared against the 60-year cycles of the SCMSS (dash curve in Figure 5B). The latter curve is shifted by +5 years to synchronize its maximum with the alignment of Jupiter and Saturn that occurred in 2000, and it is rescaled to fit the amplitude of the 60-year temperature oscillation. A remarkable correspondence is found. Figure 10A suggests that it is possible to reconstruct with a good accuracy this multidecadal climate oscillation by using the 60-year periodic component of the SCMSS with an opportune time-shift of a few years.

\begin{figure}[t!]
\includegraphics[angle=0,width=21pc]{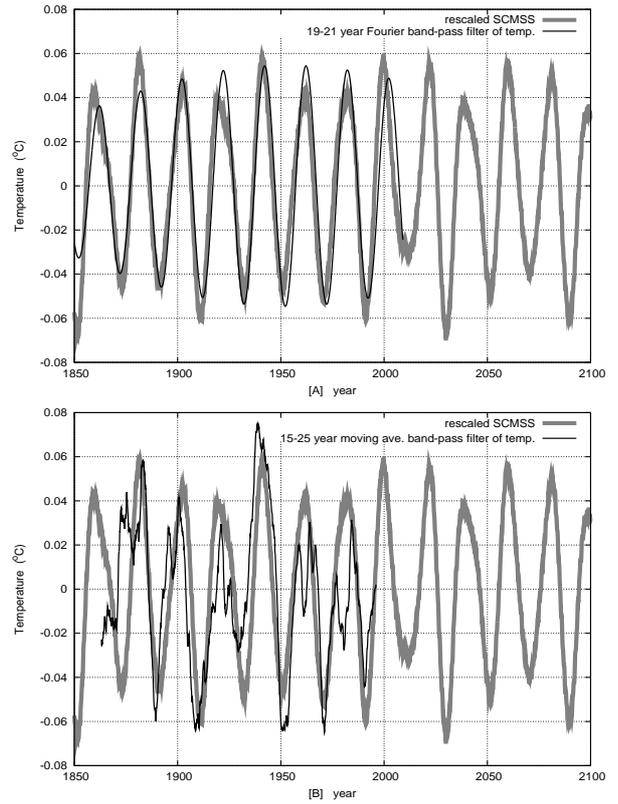}
\caption{Rescaled SCMSS against two alternative pass-band filtered records of the temperature around its two decadal oscillation. The figures clearly indicate a strong coherence between the two records.}
\end{figure}

The existence of a 60-odd year cycle in the temperature is further proven in Figure 10B. Here, the temperature residual depicted in Figure 10A is smoothed and plotted against itself with a time-shift of 61.5 years, which is the period found for this record (see Table 1). This figure shows that there exists an almost perfect correspondence between the 1880-1940 and 1940-2000 periods. These cycles have a peak-to-trough amplitude of about 0.30-0.35 $^oC$, which is about 0.30 $^oC$ from 1970 to 2000. The cross correlation between the 1880-1940 and 1940-2000 cycle periods shown in Figure 10B  gives $r \approx 0.8$, which indicates that the two periods are correlated with a probability $P>99.9\%$.

Also two smaller peaks   are clearly visible and perfectly correspond in the two superimposed records just before 1900 and 1960. These peaks are indicated with the letter `Y' in Figure 10B. These smaller cycles are due to the 20-year and 30-year periodic modulations of the temperature.

The evident strong symmetry between the 1880-1940 and 1940-2000 periods  indicates that this 60-year cycle and other shorter cycles are naturally produced. In fact, anthropogenic emissions do not show any symmetric 60-year cycle before and after the 1940s (see the figures reporting the climate forcings in Hansen \emph{et al.} [2007]).  Thus, it is very likely that at least 0.30  $^oC$ warming from 1970 to 2000 was induced by a 60-year natural cycle during its warming phase.

 Figure 10B also shows that the period 1880-1940 can be used to forecast the major climate oscillations of the period 1940-2000, or vice versa. On the basis of this finding, if this 60-year cycle repeats in the future as it did since 1650, as Figure 2 would suggest,  Figure 10B can also be used to forecast a large 60-year climate cycle until 2060 with a peak-to-trough amplitude  of at least 0.30  $^oC$. A minimum in the 2030-2040 and a new maximum in 2060 may occur. A smaller temperature peak `Y' around 2020-2022 similar to those occurred around 1900 and 1960 is expected as well.

Figure 11 shows the global surface temperature record processed with two band pass filters centered in the two-decadal cycle band. The peak-to-trough amplitude of these cycles is about 0.1 $^oC$. The curve is directly compared against  the SCMSS record(Figure 5B), which has been opportunely rescaled to reproduce the amplitude of these temperature cycles. In this case, no temporal-shift is applied, and the two curves correspond almost perfectly for most of the analyzed period. Small differences may be due to some errors in the data, pass band filter limitations, and some chaotic oscillations around limit cycles. Figure 11 suggests that it is possible to reconstruct with a good accuracy the bi-decadal climate oscillations by using the 20-year period component of the SCMSS. Perhaps, a better agreement can be obtained by taking into account also other natural cycles, such as the $22\pm2$ year solar cycle, but this is left to another study.

Figure 12 compares the global surface temperature record with two reconstructions based on the SCMSS records. These reconstructions are made of the superposition of the 20-year and the 60-year SCMSS curves depicted in Figures 10A and 11 with and without the original quadratic fit of the temperature, respectively. Figure 12B shows a reconstruction that assumes that the temperature presents three natural cycles with periods of 20, 30 and 60 years as found in the SCMSS records.  More optimized models are left to another study.

 Figures 12A and 12B clearly show that all warming and cooling  periods observed in the temperature are almost perfectly reproduced by the phenomenological celestial model: the warming observed from 1860 to 1880, from 1910 to 1940 and from 1970 to 2000; and the cooling from 1880 to 1910, from 1940 to 1970 and since 2000. Also the small warming in 1900 and 1960 is well reproduced.

 An important result of the model is that at least 60\%, that is, 0.3 $^oC$ out of the 0.5 $^oC$ global warming observed from 1970 to 2000 has been induced by the combined effect of the 20 and 60-year natural climate oscillations. In fact, both cycles had a minimum in the 1970 and a maximum in 2000. If at least 60\% of the warming observed since 1970 has been natural, humans have contributed no more than 40\% of the observed warming. This estimate should be compared with the IPCC's estimate that 100\% of the warming observed since 1970 is anthropogenic. Therefore, the climate sensitivity to anthropogenic forcing has been severely overestimated by the IPCC by a large factor.

\begin{figure} [t!]
\includegraphics[angle=0,width=21pc]{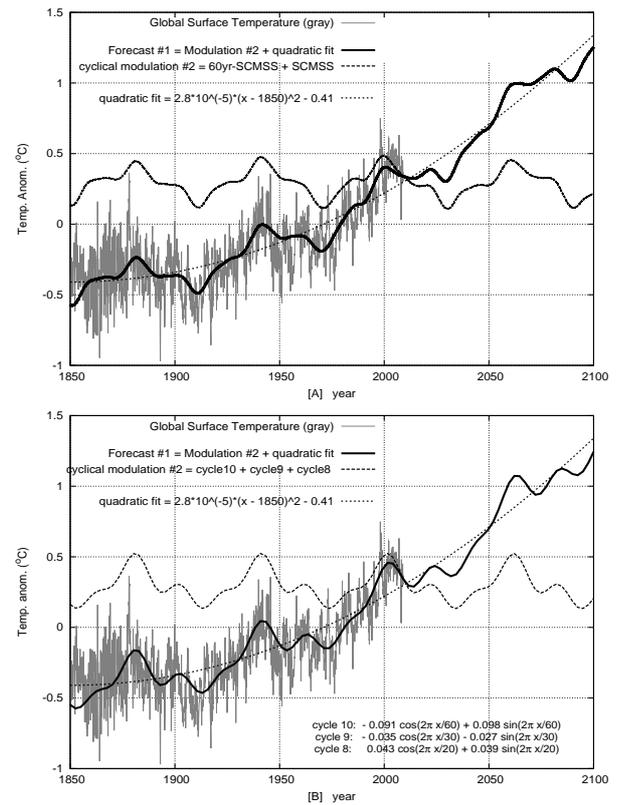}
\caption{ [A] Global temperature record (grey) and temperature reconstruction and forecast based on a SCMSS  model that uses only the 20 and 60 year period cycles (black). [B] Global temperature record (grey) and optimized temperature reconstruction and forecasts based on a SCMSS model that uses the 20, 30 and 60-year  cycles  (black). The dash horizontal curves \#2 highlight the 60-year cyclical modulation reconstructed by the SCMSS model without the secular trend.}
\end{figure}

About the 21$^{st}$ century,  scenario \#1 assumes that the temperature continues to oscillate around the  quadratic fit curve of the 1850-2009 temperature interval.  Curve \#2 shows the 60-year oscillating pattern that the model reconstructs.

If the global temperature continues to rise with the same  acceleration observed during the period 1850-2009, in 2100 the global temperature will be little bit less than 1 $^oC$ warmer than in 2009. This estimate is about three times smaller than the average projection of the IPCC [2007]. However, the meaning of the quadratic fit forecast should not be mistaken: indeed, alternative fitting functions can be adopted, they would equally well fit the data from 1850 to 2009 but may diverge during the 21$^{st}$ century. The curve depicted in the figure just suggests that if an underlying warming trend continues for the next few decades, the cooling phase of the 60-year cycle may balance the underlying upward trend. Consequently, the temperature can remain almost stable until 2030-2040. The underlying warming trend can be due to both natural and anthropogenic influences.

However,   the solar activity presents bi-secular and millennial cycles and during the last decades has reached its secular maximum. These longer cycles just started to decrease, as also Figure 5A would suggest. Perhaps, the secular component of the solar activity may continue to decrease during the 21$^{st}$ century because of its longer  cycles.

An imminent relatively long period of low solar activity can be expected on the basis that the latest solar cycle (cycle \#23) lasted from 1996 to 2009, and its length was about 13 years instead of the traditional 11 years. The only known solar cycle of comparable length (after the Maunder Minimum) occurred just at the beginning of the Dalton solar minimum (cycle \#4, 1784-1797) that lasted from about 1790 to 1830. Indeed, the last four solar cycles (\#20-23) exhibit a similarity with the four solar cycles (\#1-4) occurred just before the Dalton minimum (Scafetta, 2010). The solar Dalton minimum induced a little ice age that lasted 30-40 years.

If the secular component of the solar activity decreases during the 21$^{st}$ century, it will contribute to  a further cooling of the planet. Thus, the significant warming of the Earth predicted by the IPCC [2007] and by other AGWT advocates [Rockstr\"om J, \emph{et al.} (2009); Solomon \emph{et al.}, 2009] to occur in the following decades is unlikely. In fact, the above findings clearly suggests that the IPCC has used climate models that greatly overestimate the climate sensitivity to anthropogenic GHG increases. Therefore, the IPCC's projections for the 21$^{st}$ century are not credible.

\section{Possible physical mechanisms}

The planets, in particular Jupiter and Saturn, with their movement around the Sun give origin to large gravitational and magnetic oscillations that cause the solar system to vibrate. These vibrations have the same frequencies of the planetary orbits. The vibrations of the solar system can be directly or indirectly felt by the climate system and can cause it to oscillate with those same frequencies.

More specific physical mechanisms involved in the process include gravitational tidal forces, spin orbit transfer phenomena and magnetic perturbations (the jovian planets have large magnetic fields that interact with the solar plasma and with the magnetic field of the Earth).  These gravitational and magnetic forces act as external forcings of the solar dynamo, of the solar wind and of the Earth-Moon system and may modulate both solar dynamics and, directly or indirectly, through the Sun, the climate of the Earth.

Hung [2007] showed that an analysis of solar flare and  sunspot records  reveals a complex relation between the solar activity and the planetary gravitational tides, or at least the planetary position. Twenty-five of the thirty-eight largest known solar flares were observed to start when one or more tide-producing planets (Mercury, Venus, Earth, and Jupiter) were either nearly above the event positions ($<10^o$ longitude), or at the opposing side of the Sun. Hung [2007]  showed that the 11-year solar cycle is well synchronized with the alignment  of Venus, Earth and Jupiter. The sunspot  cycle also presents a  bi-modality with periods that oscillate between   10 and 12 years, that is between the opposition-synodic  period of Jupiter and Saturn and the period of Jupiter, respectively [Wilson, 1987]. Two large temperature cycles (\#5 and \#6) are present within this spectral range. Ogurtsov \emph{et al.} [2002] found evidences for  a 60-64 year period in $^{10}Be$, $^{14}C$ and Wolf number over the past 1000 years. Ogurtsov \emph{et al.} found 45-year cycles,  85-year cycles plus bi-secular cycles in the solar records. These findings indicate that Jupiter, Saturn, Uranus and Neptune  modulating solar  solar dynamics.

In addition, the oscillations of the magnetic field of the solar system induced by the motion of the planets (in particular Jupiter and Saturn) can influence solar plasma and solar wind. Solar wind modulates the terrestrial ionosphere that can influence the global atmospheric electric circuit. The latter effects the cloud formation and, therefore, the global climate (Tinsley, 2008).

There are other mechanisms that may link the  climate to the  motion of the planets: 1) The planets can drive solar variability and then a varying Sun can drive the climate oscillations via several amplification mechanisms and feedbacks, including, for example, a modulation of the cloud cover through the cosmic ray flux modulation [Kirkby, 2007; Svensmark \emph{et al.}, 2009]; 2) The climate can be directly influenced by the movement of the planets because their gravity, as well as their magnetic fields, act on the Earth and on the Earth-Moon system as well. In fact, the temperature records present a lunar cycle at 9.1-years. In addition, the Earth oscillates within the varying gravitational and magnetic fields generated by the Sun, the Moon and the other planets and feels the gradients of these forces.

\begin{figure} [t!]
\includegraphics[angle=-90,width=21pc]{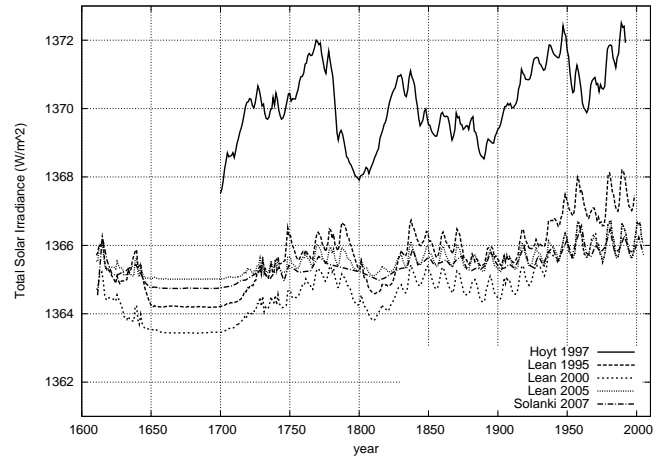}
\caption{Several proposed total solar irradiance (TSI) proxy reconstructions. [From top to bottom: Hoyt and Schatten, 1997; Lean \emph{et al.}, 1995; Lean 2000; Wang \emph{et al.}, 2005; Krivova \emph{et al.}, 2007].}
\end{figure}

If solar activity presents a quasi 60-year cycle [Ogurtsov \emph{et al.}, 2002], then some of the currently proposed total solar irradiance reconstructions, which are   shown in Figure 13, are partially erroneous. In fact, these TSI proxy reconstructions are quite different from each other: some   present a TSI peak around 1960 and a constant trend since then, but the TSI reconstruction   proposed by Hoyt and Schatten [1997] increases from 1910 to 1940, decreases from 1940 to 1970 and increases after 1970.  Hoyt and Schatten's TSI reconstruction well correlates with the temperature records during the last century [Soon, 2009].

An increase of the solar activity from 1970 to 2000 would be also supported by the ACRIM total solar irradiance (TSI) satellite composite that faithfully reproduces the satellite observations [Willson and Mordvinov, 2003; Scafetta and Willson, 2009], but not by the PMOD composite [Fr\"ohlich  and   Lean, 1998] which is the TSI record preferred  by the IPCC [2007]. From 1880 to 1910 TSI  may have been almost stable or slightly decreasing as the sunspot record did. The sunspot record has been used in several TSI proxy  reconstructions depicted in   Figure 13. Thus, it is very likely that solar activity presents a quasi-60 year modulation, although this pattern does not clearly appear in some TSI proxy reconstructions.

An additional evidence for a link between the planetary orbits and the climate is given by the length of the day (LOD) of the Earth, which presents a 60-year cycle [Stephenson and Morrison, 1995; Roberts \emph{et al.}, 2007; Mazzarella, 2007]. This cycle is almost in phase with the 60-year planetary cycle: see Figure 14. Indeed, the Earth is spinning and moving within oscillating gravitational and magnetic fields. It is possible that the Earth feels these forces as gravitational and magnetic torques that make the LOD  oscillate with those same frequencies.

It is not clear whether LOD drives the climate oscillations or vice versa. Probably there is some feedback loop that depends on the analyzed time scale. In any case, it is more likely that on a multi-decadal time scale astronomical forces drive the LOD. In fact, according to some authors [Klyashtorin, 2001; Klyashtorin and Lyubushin,  2007, 2009; Mazzarella, 2007, 2008; Sidorenkov and  Wilson, 2009]  the LOD 60-year cycle precedes the climate changes and can be used to forecast them. Moreover, there are close correlations between the PDO and the decades-long variations in the LOD, variations in the rate of the westward drift of the geomagnetic eccentric dipole, and variations in some key climate parameters such as anomalies in the type of the atmospheric circulation, the hemisphere-averaged air temperature, the increments of the Antarctic and Greenland ice sheet masses.
These results imply that the climate can be partially driven by mechanical forces such as gravitational and magnetic torques, not just radiative forces as supposed by the IPCC.  Interestingly, by using the LOD record in 2001 Klyashtorin [2001] predicted a slight cooling of the temperature that is what has been observed since 2002, while the IPCC [2001,2007] climate model projections based on GHGs  have predicted a warming [Rockstr\"om \emph{et al.}, 2009; Solomon \emph{et al.}, 2009] that did not occur [Note \#1].

The objection that the physical forces generated by the planets on the Sun and  on the Earth are \emph{small} and, therefore, it is unreasonable to believe that the planets may play any role in modulating solar and terrestrial climatic oscillations,  cannot be a conclusive argument. The above empirical findings do suggest that such claims should be questioned: evidently, physical mechanisms may exist also even when they are not understood yet. For example, Hung [2007] noticed that the tidal forces acting on the Sun could be large once all their temporal and spatial properties   are taken into account.

\begin{figure}[t!]
\includegraphics[angle=-90,width=21pc]{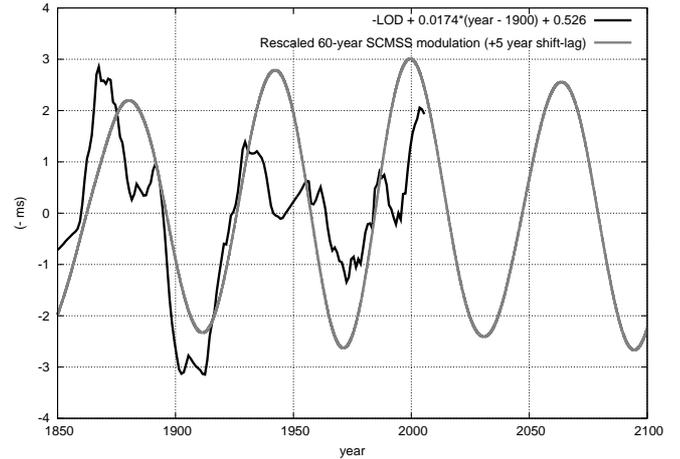}
\caption{Length of the day (LOD) (black)  against the 60-year modulation of the SCMSS (gray), which is related to the combined orbit of Jupiter and Saturn. The LOD is inverted and detrended of its linear trend, while the SCMSS is shifted by +5 years and opportunely rescaled for visual comparison. The correlation between the two records is evident.  }
\end{figure}

The objection that the Sun is in free fall and that an observable such as the SCMSS herein adopted cannot have any physical outcome is not valid. Here, we have used the SCMSS   as a proxy to calculate the frequencies of the solar system vibrations, which is perfectly appropriate. Any physical effect would be induced by tidal and magnetic forces. These forces vary with the same frequencies of the  SCMSS, or of any other function of the orbits of the planets. For example, the tidal frequencies can be calculated with the SCMSS record.

  The objection that the magnitude of the tidal forces from the jovian planets on the Sun are compatible or even smaller than those from  the terrestrial planets, and only the latter  should play a dominant role, does not take it into account that it is not only the magnitude of a tidal force that matters. The frequency of a tidal force   matters too. The tidal forces must induce a physical change on the Sun to have any physical effect and the resistance of a system such as the Sun to its own physical deformation usually works as a low pass filter and favors the low frequency forcings over the high frequency ones. This inertia likely dampens the effects of the fast varying tidal forces induced by the terrestrial planets and stresses the much slower oscillations associated to the tidal forces induced by the jovian planets. Consequently, the effect of the tidal forces induced by the jovian planets should be the dominant one. A similar reasoning applies to the forces acting on the Earth. A detailed discussion on this topic is left to another study.

  Complex systems such as the Sun and the Earth likely contain mechanisms that can amplify the effect of small external perturbations. In addition to radiative forcings and feedback mechanisms [Scafetta, 2009], periodic forcings can easily stimulate resonance and give rise to collective synchronization of coupled oscillators. Resonance, collective synchronization and feedback mechanisms amplify the effects of a weak external periodic forcing. This would be true both for the Sun's dynamics as well as for the Earth-Moon system and, ultimately, for the climate.

For example, collective synchronization of coupled oscillators, which is a very common phenomenon among complex chaotic systems, was first noted by Huygens in the 17$^{th}$  century. This phenomenon means adjustment of the rhythms of self-sustained periodic oscillators due to their weak interactions. If a system of coupled oscillators is forced by an external periodic forcing, even if this forcing is weak, it can force the oscillators of the system to synchronize and follow the frequency of the external forcing. Consequently, after a while, the entire system oscillates at the same frequencies of the input weak forcing.

In collective synchronization,   a weak periodic external forcing is not the primary source of the energy that makes the system  oscillate. The system has its own self-sustained oscillators. The external forcing simply drives the adjustment of the natural rhythms of the system by letting their own energy  flow with the same frequency of the forcing.  Thus, the external forcing just passes to the system the \emph{information} [Scafetta and West, 2008] of how it has to oscillate, not the entire energy to make it  oscillate.  The effect of a periodic external forcing, even if weak, may become macroscopic. That is, the system can mirror the frequency of the input weak forcing by means of collective synchronization of its own internal oscillators. In other words, all components of the system gradually synchronize with the external forcing. More realistically, a system chaotically oscillates around limit cycles driven by the frequencies of the external forcings.  This gives origin to a constructive interference signal in the system and, therefore, to an amplification of the effect of the input forcing signal. The properties of an elementary model of collective synchronization of coupled oscillators are discussed in the Appendix.

\section{Conclusion}

On secular, millenarian and larger time scales astronomical oscillations and solar changes drive climate variations.  Shaviv's theory [2003] can explain the large 145 Myr climate oscillations during the last 600 million years.  Milankovic's theory [1941] can explain the multi-millennial climate oscillations observed during the last 1000 kyr. Climate oscillations with periods of 2500, 1500, and 1000 years during the last 10,000 year (the Holocene) are correlated to equivalent  solar cycles that caused the Minoan, Roman, Medieval and Modern warm periods [Bond \emph{et al.}, 2001; Kerr, 2001]. Finally, several other authors found that multisecular  solar oscillations  caused bi-secular little ice ages (for example: the Sp\"orer, Maunder, Dalton minima) during the last 1000 years [for example: Eddy, 1976; Eichler \emph{et al.}, 2009; Scafetta and  West, 2007;  Scafetta, 2009, 2010].

Herein, we have found empirical evidences that the climate oscillations within the secular scale are very likely driven by astronomical cycles, too. Cycles with periods of 10-11, 12, 15, 20-22, 30 and 60 years are present in all major surface temperature records since 1850, and can be easily linked to the orbits of Jupiter and Saturn. The 11 and 22-year cycles are the well-known Schwabe and Hale solar cycles.  Other faster cycles with periods between 5 and 10 years are in common between the temperature records and the astronomical cycles. Long-term lunar cycles induce a 9.1-year cycle in the temperature records and probably other cycles, including an 18.6-year cycle in some regions [McKinnell and Crawford, 2007]. A quasi-60 year cycle has been found in numerous multi-secular climatic records, and it is even present in the traditional Chinese, Tibetan and Tamil calendars, which are arranged in major 60-year cycles.

The physical mechanisms that would explain this result are still unknown. Perhaps the four jovian planets modulate solar activity via gravitational and magnetic forces that cause tidal and angular momentum stresses on the Sun and its heliosphere. Then, a varying Sun modulates climate, which amplifies the effects of the solar input through several feedback mechanisms. This phenomenon is mostly regulated by Jupiter and Saturn, plus some important contribution from Neptune and Uranus, which modulate a bi-secular cycle with their 172 year synodic period. This interpretation is supported by the fact that the 11-year solar cycles and the solar flare occurrence appear synchronized to the tides generated on the Sun by Venus, Earth and Jupiter [Hung, 2007]. Moreover, a 60-year cycle and other planetary cycles have been found in millennial solar records [Ogurtsov \emph{et al.}, 2002] and in the number of middle latitude auroras [Komitov, 2009].

Alternatively, the planets are directly influencing the Earth's climate by modulating the orbital parameters of the Earth-Moon system and of the Earth. Orbital parameters can modulate the Earth's angular momentum via gravitational tides and magnetic forces. Then, these orbital oscillations are amplified by the climate system through synchronization of its natural oscillators. This interpretation is supported by the fact that the temperature records contain a clear 9.1-year cycle, which is associated to some long-term lunar tidal cycles. However, the climatic influence of the Moon may be more subtle because several planetary cycles are also found in the Earth-Moon system.

The astronomical forcings may be modulating the length of the day (LOD). LOD presents a 60-year cycle that anticipates the 60-year temperature cycle [Klyashtorin 2001; Klyashtorin and Lyubushin,  2007, 2009; Mazzarella, 2007, 2008; Sidorenkov and Wilson, 2009]. A LOD change  can drive the ocean oscillations by exerting some pressure on the ocean floor and by modifying the Coriolis' forces. In particular, the large ocean oscillations such as the AMO and PDO oscillations are likely driven by astronomical oscillations.

The results herein found show   that the climate oscillations are driven by multiple astronomical mechanisms. Indeed, the planets with their movement cause the entire solar system to vibrate with a set of frequencies that are closely related to the orbital periods of the planets. The wobbling of the Sun around the center of mass of the solar system is just the clearest manifestation of these solar system vibrations and has been used herein just as a proxy for studying those vibrations. The Sun, the Earth-Moon system and the Earth feel these oscillations, and it is reasonable that the internal physical processes of the Earth and the Sun synchronize to them.

It is evident that we can still infer, by means of a detailed data analysis, that the solar system likely induces the climate oscillations, although the actual mechanisms that explain the observed climate oscillations are still unknown. If the true climate mechanisms were already known and well understood, the general circulation climate models would properly reproduce the climate oscillations. However, we found that this is not the case. For example, we showed that the GISS ModelE fails to reproduce the climate oscillations at multiple time scales, including the large 60-year cycle. This failure is common to all climate models adopted by the IPCC [2007] as it is evident in their figures 9.5 and SPM.5 that show the multi-model global average simulation of surface warming. This failure indicates that the models on which the IPCC's claims are based are still incomplete and possibly flawed.

 The existence of a 60-year natural cycle in the climate system, which is clearly proven in multiple studies and herein in Figures 2, 6, 10 and 12, indicates that the AGWT promoted by the IPCC [2007], which claims that 100\% of the global warming observed since 1970 is anthropogenic, is erroneous.   In fact, since 1970 a global warming of about 0.5 $^oC$ has been observed. However, from 1970 to 2000 the 60-year natural cycle was in his warming phase and has contributed no less than 0.3 $^oC$ of the observed 0.5 $^oC$ warming, as Figure 10B shows. Thus, at least 60\% of the observed warming since 1970 has been naturally induced.   This leaves less than 40\% of the observed warming to human emissions.
   Consequently, the current climate models, by failing to simulate the observed quasi-60 year temperature cycle, have significantly overestimated the climate sensitivity to anthropogenic GHG emissions by likely a factor of three. Moreover, the upward trend observed in the temperature data since 1900 may be partially due to land change use, uncorrected urban heat island effects [McKitrick and Michaels, 2007; McKitrick, 2010] and to the bi-secular and millennial solar cycles that reached their maxima during the last decades [Bond \emph{et al.}, 2001; Kerr, 2001; Eichler \emph{et al.}, 2009; Scafetta, 2010].

 Solomon \emph{et al.} [2010] recently acknowledged that stratospheric water vapor, not just anthropogenic GHGs, is a very important climate driver of the decadal global surface climate change.  Solomon \emph{et al.} estimated that stratospheric water vapor has largely contributed both to the warming observed from 1980-2000 (by 30\%) and to the slight cooling observed after 2000 (by 25\%). This study reinforces that climate change is more complex than just a response to added $CO_2$ and a few other anthropogenic GHGs. The causes of stratospheric water vapor variation are not understood yet. Perhaps, stratospheric water vapor is driven by UV solar irradiance variations through ozone modulation, and works as a climate feedback to solar variation [Stuber \emph{ et al.}, 2001]. Thus, Solomon's finding would partially support the findings of this paper and those of Scafetta and West [2005, 2007] and Scafetta [2009]. The latter studies found a significant natural and solar contribution to the warming from 1970-2000 and to the cooling afterward.

  A detailed reconstruction of the climate oscillations suggests that a   model based on celestial oscillations, as shown in Figure 12, would largely outperform current general circulation climate models, such as the GISS ModelE, in reconstructing the climate oscillations. The planetary model would also be  more accurate in forecasting climate changes during the next few decades. Over this time, the global surface temperature will likely remain approximately steady, or actually cool.

 In conclusion, data analysis indicates that current general circulation climate models are missing fundamental mechanisms that have their physical origin and ultimate justification in astronomical phenomena, and in interplanetary and solar-planetary interaction physics.

\section*{Appendix: Collective synchronization  of coupled oscillators}

Herein we  briefly discuss some mathematical properties of the phenomenon known as  collective synchronization  of coupled oscillators, which was first noticed by Huygens in 1657 [Strogatz, 2009]. We discuss a simple forced Kuramoto model [Kuramoto, 1984].

Collective synchronization is a common physical mechanism that can greatly amplify the effect of a periodic input forcing. Synchronization mechanisms can explain how small periodic extraterrestrial forcings  can be mirrored by the  climate system and contribute to a terrestrial amplification of a weak external periodic forcing. Synchronization in climate  has been observed in the results herein obtained and by other authors [Tsonis \emph{et al.}, 2007].

This simple model assumes that there are N coupled oscillators, which in this example would represent the natural internal oscillators of the climate. Each oscillator by alone would produce a signal equal to $X_i(t)=A_i ~\sin[\theta_i(t)]$ (for example a temperature)  characterized  by a given natural frequency $\omega_i$. Its evolution is described by the phase function $\theta_i(t)$ for $i=1,2,\dots,N$. Without any coupling each oscillator varies with its own natural frequency: $\theta_i(t)=\omega_i t+\theta_i(0)$.

These $N$ oscillators form a coupled network and the coupling coefficients between the oscillator $i$ and the oscillator $j$ is $K_{ij}$. We assume that there exists an external periodic forcing characterized by a frequency $\omega$. Its phase evolves in time as $\theta (t)=\omega t$. This external forcing is coupled to each oscillator by an appropriate coupling coefficient $K_{i}$.

The entire system is made of a set of coupled equations of the type:
\begin{equation}\label{eq1}
    \frac{d\theta_i}{dt}=\omega_i+\frac{1}{N}\sum_{j=1}^N K_{ij}~ sin(\theta_j-\theta_i)+K_{i}~ sin(\omega t-\theta_i),
\end{equation}
where $i=1,2,\dots,N$. The system can be run by imposing at $t=0$ random initial phases $\theta_i(t=0)$ and random internal frequencies $\omega_i$.

In the absence of the external forcing ($K_i=0$), all oscillators of the system will gradually synchronize. The synchronization frequency is called \emph{mean field frequency} [Strogatz, 2009] and it is a characteristic of the network system. The mean field frequency is given by the average of all frequencies $\omega_i$ if all coupling coefficients $K_{ij}$ are equal.

The purpose of this exercise is to show what happens if the system is forced by a weak external periodic forcing. In the simulation, we assume that the coupling coefficient among the oscillators of the system is $K_{ij}=3$ and the coupling of the forcing with each oscillator is $K_i=0.3$. In this specific simulation, a system of 20 coupled oscillators is used and their mean field frequency is $\bar{\omega}=0.41$. The frequency of the forcing is set to $\omega=0.7$.

 Figure 15 shows the result of the simulation. The 20 oscillators start from random phases and each has its own frequency. The figure shows that the oscillators rapidly synchronize first to the mean field frequency mode because of the strong coupling between them. Then, all of them synchronize to the external weak forcing. At this point all oscillators move in phase, and with the same frequency of the input forcing.

\begin{figure} [t!]
\includegraphics[angle=-90,width=21pc]{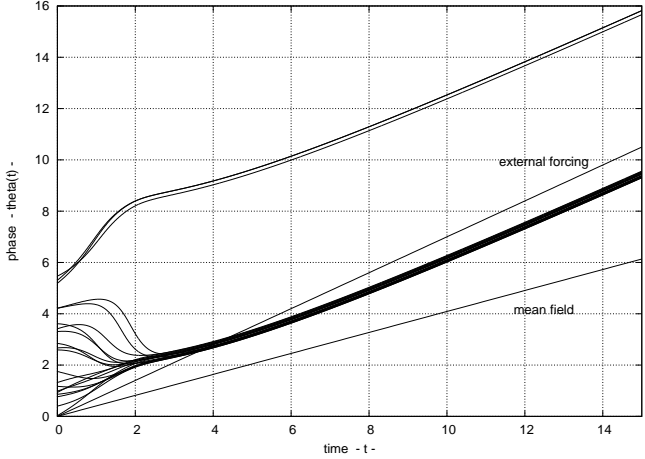}
\caption{Collective synchronization of 20 coupled oscillators [Strogatz, 2009]. A simple forced Kuramoto model Eq. (\ref{eq1}). The coupled oscillators of the system first synchronize to the internal mean field frequency $\bar{\omega}=0.41$, and later to the external input forcing frequency $\omega=0.7$, as indicated by the slopes of the curves.}
\end{figure}

Without synchronization, the total average signal emerging from the system would be given by:
  \begin{equation}\label{}
   X(t)=\frac{1}{N}\sum_{i=1}^N X_i(t)=\frac{1}{N}\sum_{i=1}^N A_i ~\sin[\theta_i(t)].
  \end{equation}
  The amplitude of $X(t)$ is expected to be relatively small and irregular because the signals $X_i(t)$ are out of phase and each  oscillates with its own frequency.

  With synchronization, the total average signal emerging from the system is given by:
  \begin{equation}\label{}
   X(t)=\frac{1}{N}\sum_{i=1}^N X_i(t)=\frac{1}{N}\left[\sum_{i=1}^N A_i\right] \sin[\omega t+\alpha],
  \end{equation}
   where $\alpha$ is a given common phase and $\omega$ is the frequency of the forcing, which  has been \emph{adopted} by the N oscillators of the system. Thus, after synchronization the signal $X(t)$ has a large amplitude and oscillates with the same frequency of the input forcing signal. We can say that through synchronization of its oscillators, the system has significantly amplified the effect of the input forcing signal by mirroring it in its own dynamics.

Note that the above model is bi-stable in the sense that there are two possible synchronization frequencies: the external forcing frequency and the internal mean field frequency. The system may switch from one mode to the other according to the strength of the couplings among the oscillators of the system and with the external forcing. This can give origin to a chaotic variability in the dynamics of the system, in particular if the strength of the couplings and of the external forcing changes in time.

In any case, the suggested model is just a simple mathematical prototype that suggests how the climate may synchronize with weak astronomical periodic forcings by just adjusting, through synchronization, the frequency modes of its own numerous internal subsystems in such a way to let them  mirror the oscillations of the input forcing. This synchronization mechanism  acts in addition and together with other more direct mechanisms such as irradiance forcing and cloud modulation via cosmic ray flux [Kirkby, 2007; Svensmark \emph{et al.}, 2009],  and contributes to magnifying the climatic effect of a weak  astronomical periodic forcing.

\section*{The bibliography}

\onecolumn

\begin{table}
  \centering
\begin{tabular}{|c|c|c|c|c|c|c|c|c|c|}
  \hline
 P		&	G	&	GN	&	GS	&	L	&	LN	&	LS	&	O	&	ON	 &	OS				 \\\hline
1	&	5.99	&	5.99	&	6.03	&	6.1	&	6.08	&	5.98	 &	5.98	&	5.98	 &	 6.03		 \\\hline
2	&	6.45	&	n	&	6.48	&	6.59	&	6.76	&	6.47	 &	6.49	&	n	&	 6.5		 \\\hline
3	&	7.5	&	7.45	&	7.44	&	7.5	&	7.39	&	7.54	&	 7.53	&	7.53	&	 7.43		 \\\hline
4	&	8.35	&	n	&	8.05	&	8.4	&	8.41	&	8.37	&	 8.2	&	n	&	7.83		 \\\hline
M	&	9.01	&	9.05	&	9.15	&	9.05	&	9.03	&	 9.1	&	9.08	&	9.05	 &	 9.14			 \\\hline
5	&	10.43	&	10.59	&	10.43	&	10.14	&	9.8	&	10.29	 &	10.46	&	10.6	 &	 10.43		 \\\hline
6	&	12.3	&	11.78	&	n	&	11.5	&	11.56	&	11.47	 &	12.56	&	12.18	 &	 12.8		 \\\hline
7	&	14.8	&	15	&	14.9	&	14.8	&	15.4	&	14.6	 &	14.9	&	14.9	 &	 14.8		 \\\hline
8	&	20.9	&	21.1	&	20.2	&	21.4	&	21	&	21.8	 &	21.3	&	21.5	 &	 20		 \\\hline
9	&	31.6	&	30.4	&	29.2	&	28.2	&	29.8	&	 29.3	&	32.9	&	32	 &	 27.9		 \\\hline
10	&	61.5	&	61.6	&	59.7	&	67.2	&	67.4	&	 62.2	&	60.6	&	 60.6	 &	 58.7		 \\\hline	
\end{tabular}
  \caption{The period in years of the eleven spectral cycles (P) indicated in Figure 6B for each of the nine temperature records:  global temperature (G);	 northern hemisphere (GN); southern hemisphere 	 (GS); global land temperature	 (L); northern hemisphere land (LN); southern hemisphere land (LS); 	 global ocean temperature	 (O); northern hemisphere ocean (ON); southern hemisphere ocean 	(OS).  These peaks have a 99\% confidence interval. The letter `n' indicates that no cycle is easily recognizable in that frequency band. The letter ``M'' indicates the large cycle that does not appear in the SCMSS record. This cycle is linked to  lunar cycles: see Figure 8.}\label{}
\end{table}

\begin{table}
  \centering
\begin{tabular}{|c|c|c|c|}
  \hline
P	&	Average Temp.			&	SCMSS/Sun/Moon			&	GISS ModelE			\\\hline
1	&	6.02	$\pm$	0.13	&	5.91	$\pm$	0.07	&	6.05	 $\pm$	0.07	 \\\hline
2	&	6.53	$\pm$	0.17	&	6.56	$\pm$	0.07	&	6.73	 $\pm$	0.07	 \\\hline
3	&	7.48	$\pm$	0.16	&	7.52	$\pm$	0.08	&	7.22	 $\pm$	0.08	 \\\hline
4	&	8.23	$\pm$	0.27	&	8.07	$\pm$	0.08	&	7.83	 $\pm$	0.08	 \\\hline
M	&	9.07	$\pm$	0.19	&	(9.1	$\pm$	0.1)	&	8.94	 $\pm$	0.1	\\\hline
5	&	10.35	$\pm$	0.32	&	9.84	$\pm$	0.12	&	9.9	 $\pm$	0.2	\\
	&		&	(11	$\pm$	1)	&		\\\hline
6	&	12.02	$\pm$	0.89	&	11.8	$\pm$	0.18	&	13.4	 $\pm$	0.3	\\\hline
7	&	14.9	$\pm$	0.92	&	14.2	$\pm$	0.4	&	15.5	 $\pm$	0.4	\\\hline
8	&	21.02	$\pm$	1.39	&	20.2	$\pm$	0.7	&	20.4	 $\pm$	0.5	\\
	&			&	(22	$\pm$	2)	&		\\\hline
9	&	30.14	$\pm$	2.5	&	30.5	$\pm$	1	&	25.3	$\pm$	 0.7	\\\hline
10	&	62.17	$\pm$	4.85	&	59.9	$\pm$	2	&	72	$\pm$	 5	\\\hline
	&					&$\tilde{\chi}_O^2\leq0.38$&				 $\tilde{\chi}_O^2=1.4$			 \\
	&					&$P_{11}(\tilde{\chi}^2 \geq \tilde{\chi}_O^2)\geq 96\%$&				 $P_{11}(\tilde{\chi}^2 \geq \tilde{\chi}_O^2)\approx 16\%$			 \\\hline
\end{tabular}
  \caption{First column: average of the eleven temperature spectral periods (P) reported in Table 1. Second column: SCMSS spectral   periods   shown in Figure 6. Third column:  GISS ModelE spectral periods,   Figure 7. The error bars are calculated by taking into account the standard deviation among the frequencies and the width of the peaks at half high. The latter is between 2\% and 6\% of the peak value that corresponds to approximately 99\% confidence interval.  The row indicated by the letter ``M'' refers to the lunar cycle: see Figure 8. The rows ``5'' and ``8'' under the column SCMSS also report the Schwabe and Hale solar cycles (in parentheses), which are shown in Figure 6.  The last row reports the reduced $\tilde{\chi}_O^2$ between the temperature and the SCMSS, and between the temperature and the GISS ModelE sets of frequency peaks, respectively.}\label{}
\end{table}

.\newpage

\end{document}